
\magnification=1200
\overfullrule=0pt
\baselineskip=15pt
\font\mfs=cmr9
\font\rmb=cmr9 scaled \magstep 1
\font\rma=cmr9 scaled \magstep 2
\font\rmm=cmr9 scaled \magstep 3
\def\citem{\par\noindent\hangindent=0.5cm \hangafter=1  }

\def\R{\rm I\kern-.18em R}
\def\Z{\rm Z\kern-.4em Z}
\def\E{\rm I\kern-.18em E}
\def\nor#1{\vert #1 \vert}
\def\hn{$\;{\rm h}^{-1}$}
\def\etal{{\it et al. }}

\topskip 6 true cm
\pageno=0

\centerline{\rmm ROBUST MORPHOLOGICAL MEASURES}
\bigskip
\centerline{\rmm FOR LARGE-SCALE STRUCTURE IN THE UNIVERSE}
\bigskip\bigskip
\centerline{\rmm by}
\bigskip\bigskip
\centerline{\rmm K.R. Mecke$^{1}$, T. Buchert$^{2}$, H. Wagner$^{1}$}
\vskip 1 true cm
\centerline{\rma $^{1}$Sektion Physik der Universit\"at M\"unchen}
\smallskip
\centerline{\rma Theresienstr. 37}
\smallskip
\centerline{\rma 80333 M\"unchen}
\smallskip
\centerline{\rma G e r m a n y}
\bigskip
\centerline{\rma $^{2}$Max-Planck-Institut f{\"u}r Astrophysik}
\smallskip
\centerline{\rma Postfach 1523}
\smallskip
\centerline{\rma 85740 Garching bei M\"unchen}
\smallskip
\centerline{\rma G e r m a n y}
\vskip 2 true cm
\centerline{\rmb submitted to {\it Astron. Astrophys., Main
Journal}}

\vfill\eject

\centerline{\rmm Robust morphological measures}
\smallskip
\centerline{\rmm for large-scale structure in the Universe}

\bigskip\bigskip
\centerline{\rma by}
\bigskip\bigskip
\centerline{\rma K.R. Mecke, T. Buchert, H. Wagner}
\vskip 1.3 true cm
\noindent
{\mfs
{\narrower
{\rma Summary:}
We propose a novel method for the description of spatial patterns
formed by a coverage of point sets representing galaxy samples.
This method is based on a complete family of morphological measures
known as Minkowski functionals, which includes the topological
Euler characteristic and geometric descriptors to specify the content,
shape and connectivity of spatial sets.

\noindent
The method is numerically robust even for small samples, independent
of statistical assumptions, and yields global as well as local
morphological information.
We illustrate the method by applying it to
a Poisson process, a `double-Poisson'
process, and to the Abell catalogue of galaxy clusters.

}}

\vfill\eject
\topskip= -6 true cm

\noindent{\rmm 1. Introduction}
\bigskip\bigskip
\noindent
Statistical measures
provide important tools for the comparison of inhomogeneous
cosmological models with observations of large-scale structure in the
Universe. The discriminative power of this
comparison depends chiefly on the statistical measure used.
In the cosmological community the most frequently used measure
was and still is the two-point correlation function of a point process
such as the distribution of galaxies (e.g., Peebles 1980 and references
therein, Davis \& Peebles 1983),
or clusters of galaxies (e.g., Bahcall \& Soneira 1983).
This measure appears to be crude: if we consider a cosmological
density field, then the two-point correlation function only embodies
information contained in the power spectrum of this field;
it does not provide
any information on the distribution of the phases of the Fourier
components of the field, and is limited to spatial regions which are
well below
currently observed large-scale structures such as, e.g., the ``Great
Wall'' (Geller \& Huchra 1989).
As a consequence, completely different spatial patterns
could display the same two-point correlation function
(Mart\'\i nez \etal 1990), i.e., no direct conclusions about the
morphology of large-scale structure can be drawn.
However, in general, the characteristics of the
two-point correlation function as a small-scale measure are affected
by the possible existence of
large-scale inhomogeneities in an uncontrollable way.
Moreover, they could depend strongly
on `biasing' assumptions about the relation between dark and luminous
matter (Mart\'\i nez \etal 1993, Buchert \& Mart\'\i nez 1993).

\medskip
Efforts to go beyond the two-point correlation function comprise
the consideration of higher-order correlations (Peebles 1980 and ref.
therein), integrated unnormalized pair-counts (Mo \etal 1992),
void probability
functions (White 1979), percolation analysis (Shandarin 1983),
minimal spanning trees (Barrow \etal 1985), Voronoi foam statistics
(Icke \& van de Weygaert 1987, van de Weygaert 1991),
cell-count variance (Efstathiou \etal 1990), topological/geometrical
structure discriminators (Mo \& Buchert 1990),
and multi-fractal measures (Mart\'\i nez 1991).
However, statistical measures which are sensitive to the morphology
of structures have been investigated mostly in other fields
such as image analysis and pattern recognition
(Rosenfeld \& Kak 1976 and ref. therein, Serra 1982).

\medskip
Of those approaches which focus on global aspects
of the matter distribution, the use of topological descriptors
appears to be among the
most effective with respect to the comparison of cosmological
theories and observations. Melott and collaborators
(see the
review by Melott 1990 and ref. therein) have pursued this possibility
and have investigated methods to probe the topology of large-scale
structure by employing the Euler characteristic or the genus of
iso-density level surfaces of a density field obtained by smoothing
the observed galaxy distribution. Recently, this method was modified
by Brandenberger \etal (1993) into a {\it discrete genus statistic}.
\medskip

This approach translates the phenomenology of the large-scale
distribution of matter we observe in the Universe into a
quantitative statement of connectedness of these structures:
for different iso-density levels different topologies are realized
by the same matter distribution. For high density levels
isolated islands remain showing a {\it bubble} or {\it cluster}
topology, for low density levels a connected structure
occurs which has the topology of a {\it honeycomb} or {\it swisscheese},
intermediate
topologies are named {\it spongy}, i.e., holes and islands are
interlocking (compare Melott 1988, 1990). Also, the change of the
topology of large-scale
structure can be an evolutionary effect: while the early distribution
of QSO's exhibits a ``bubble'' topology, the evolution of
overdense structures into interconnecting sheets
of superclusters displays the opposite topology as is
seen in the evolution of pancake models typical for a {\it HDM}
(Hot-Dark-Matter) cosmogony.
\medskip
The full morphological specification of spatial patterns requires
{\it topological} as well as {\it geometrical} descriptors to
characterize not only the connectivity but also the content and shape
of figures. The aim of our paper is to point out that integral geometry
supplies a suitable family of such descriptors, known as {\it
Minkowski functionals}. In a $d$--dimensional ambient space there exist
$d+1$ of these functionals. We are primarily interested in
the case $d=3$,
where the Minkowski functionals are related with familiar measures:
covered volume, surface area, integral mean curvature, and
Euler characteristic.
\medskip
In order to apply the functionals to point-sets representing
galaxy samples, we proceed by decorating each point with a spherical
ball. The scale-dependent morphological features of this coverage
may then be explored by varying the radius of the balls.
\medskip
The Minkowski functionals are {\it global} and {\it additive} measures.
Additivity allows us to
calculate these measures effectively by summing up their
{\it local} contributions, as in the case of the
Euler characteristic. Moreover, the functionals are numerically {\it
robust} against short-scale spatial irregularities in the coverage,
which may arise via the variation of the ball radius.
Finally, a further advantage is that the computation does not rely on
statistical assumptions concerning the galaxy distribution.
\bigskip
In the next section we briefly review the mathematical background.
In Section 3 the results of sample calculations are described.
Section 4 gives a summary.

\vfill\eject

\noindent{\rmm 2. Minkowski functionals}
\bigskip\medskip

\noindent
Consider a set of points $\{x_i / i=1,\ldots,N\}$
which represent the positions of a sample of
galaxies in three-dimensional Euclidean space $\E^3$.
In order to study the global features of
this point set, we introduce a collection of neighborhoods by covering
each point with a spherical
ball $B_r(x_i)=\{x \in \E^3 / \nor{\nor{x-x_i}} \leq r\}$.
For any chosen value of the radius $r$
the union of these balls specifies an ensemble of clusters,  where two
balls are said to belong to
the same cluster if they are connected by a chain of intersecting
balls. By employing the
Minkowski functionals to measure
{\it content}, {\it shape} and {\it connectivity} of
the covering ${\cal B}(r)=\cup_{i=1}^{N} B_r(x_i)$
under variation of $r$, we arrive at
a quantitative operational definition for the
scale-dependent morphological characteristics of the
underlying spatial point process.
\medskip
Since the methods of integral geometry are perhaps not widely
known among physicists, we compile in this section some pertinent
facts (Hadwiger 1957, Santal\'o 1976, Weil 1983).
\medskip
The Minkowski functionals derive from the theory of convex sets.
They generalize curvature integrals over smooth surfaces to the case
of surfaces with singular edges and corners. In the example of the
covering ${\cal B}(r)$ such irregularities arise from the intersections
of the decorating balls.
\medskip
Let $K$ be a compact convex body in $\E^3$, with {\it regular} boundary
$\partial K \in {\cal C}^2$ and
principal radii of curvature $R_1$ and $R_2$.
The surface area $F$ of $K$ is given by
$$
F= \int_{\partial K} df = \int_{S^2}R_1R_2d\sigma \;\; ,  \eqno(1)
$$
since the area element $df$ on $\partial K$ is related with the
area element $d\sigma$ of the
spherical image of $\partial K$ under the Gaussian map by
$df=R_1R_2 d\sigma$.
\medskip
Consider the convex body $K_\epsilon$
parallel to $K$ at a distance $\epsilon$,
i.e.:
$$
K_\epsilon = \cup_{x\in K}B_\epsilon(x) \;\; .\eqno(2)
$$
Thus, the parallel body $K_{\epsilon}$ consists of all points within a
distance $\le \epsilon$ from $K$.

\noindent
The radii of curvature of $\partial K_\epsilon$
are $R_1+\epsilon$ and $R_2+\epsilon$. Therefore
$$
F_\epsilon= \int_{S^2}(R_1+\epsilon)(R_2+\epsilon)d\sigma
=F+2\cdot H \cdot \epsilon+ G\cdot \epsilon^2 \;\; , \eqno(3)
$$
where
$$
H= {1\over 2}
\int_{\partial K}\left({1 \over R_1}+{1 \over R_2}\right) df \eqno(4)
$$
is the {\it integral mean curvature} and
$$
G= \int_{\partial K}{1 \over R_1 R_2}df  \eqno(5)
$$
is the {\it integral Gaussian curvature} of $\partial K$.

\noindent
If $V$ denotes the volume of $K$ and $V_\epsilon$
that of $K_\epsilon$, we have
$$
V_\epsilon= V+\int_{0}^\epsilon F_{\epsilon'} d\epsilon' =
V+F\cdot \epsilon+ H \cdot \epsilon^2+{4 \pi \over 3} \epsilon^3 \;\; ,
\eqno(6)
$$
which is known as {\it Steiner's formula}.

More generally, Steiner's formula for convex bodies
in d-dimensional Euclidean space may
be written as
$$
V_\epsilon(K)=\sum_{\alpha=0}^d {d \choose \alpha}
W_\alpha (K) \epsilon^\alpha \;\;,\eqno(7)
$$
whereby the Minkowski functionals $W_\alpha(K)$, $\alpha=0,\ldots,d$
are defined. For $d=3$ we find by comparison
$$
W_0(K)=V(K)\;,\;\;3W_1(K)=F(K)\;,\;\;
3W_2(K)=H(K)\;,\;\;3W_3(K)=G(K)=4\pi\chi(K)\;\;,\eqno(8)
$$
with the constant $\chi(K)=1$ denoting the Euler characteristic of a
convex body.

The expression (7) for $V_\epsilon(K)$
allows us to extend the definition of the functional
$W_\alpha$ to any convex body without requiring
regularity of $\partial K$; compare Fig.1. For
example, let us take a cube $Q$ in $\E^3$ with edge length $a$.
Its parallel volume
$V_\epsilon(Q)$ may be found by inspection,
$$
V_\epsilon(Q)=a^3+6a^2 \epsilon+ 12 a {\pi \over 4} \epsilon^2+
8 {4 \pi \over 24} \epsilon^3 \;\; .\eqno(9)
$$
The $\epsilon$-terms are the contributions
from the rectangular prisms on the faces of $Q$, from
the cylindrical sectors on its edges and from
the spherical sectors located on its corners.

Furthermore, Steiner's formula also yields the
Minkowski functionals for lower dimensional
(``improper'') convex bodies embedded in $\E^d$, such as points,
straight line segments, discs etc.$\;$.

\noindent
For example, in the case of a disk $D_r$ with radius $r$ in $\E^3$
one finds
$$
V_{\epsilon}(D_r)=2\pi r^2 \epsilon + \pi^2 r \epsilon^2 +{4\pi\over3}
\epsilon^3 \;\;,\eqno(10)
$$
and thus
$$
W_0(D_r)=V(D_r)=0\;,\;3W_1(D_r)=F(D_r)=2\pi r^2\;,\;
$$
$$
3W_2(D_r)=H(D_r)=\pi^2 r\;,\;3W_3(D_r)=G(D_r)=4\pi\chi(D_r)=4\pi
\;\;.\eqno(11)
$$
We note that the two-dimensional disk enters here as a cylindrical
prism with vanishing height; therefore its ``content'' $F(D_r)=2\pi r^2$
is twice the area of a two-dimensional ball with radius $r$. If
$r\rightarrow 0$, the disk degenerates into a point $P=D_{r=0}$ with the
Euler characteristic $\chi(P)=1$ and $W_{\alpha}(P)=0,\alpha=0 \ldots
d-1$.

\bigskip
Let us now list some common general
properties of the functionals $W_\alpha : {\cal K}
\rightarrow \R$. Here, ${\cal K}$
denotes the class of closed bounded convex subsets of $\E^d$.
\medskip\noindent
$\underline{\rm Additivity:}$ If $K \in \cal K$
is dissected by a planar cut
into two convex parts such that $K=K_1 \cup K_2$, $K_1,K_2 \in \cal K$,
then
$$
W_\alpha(K_1\cup K_2)=W_\alpha(K_1)+
W_\alpha(K_2)-W_\alpha(K_1\cap K_2)\; , \eqno(12)
$$
holds for each $\alpha = 0, \ldots,d$.

\noindent
This relation is easily verified if $K=B_r(x)$ is a ball in $\E^3$
which is cut into two hemispheres $K_1$ and $K_2$. For instance, the
surface area $(4\pi r^2)$ of $K_1 \cup K_2$
is obtained by adding up the surface areas of the hemispheres
$(2\cdot(2\pi r^2+\pi r^2))$ and subtracting the area $(2\pi r^2)$
of the disk $D_r = K_1 \cap K_2$.

\medskip\noindent
$\underline{\rm Motion\ invariance:}$
Let $\cal G$ be the group of motions
(translations and rotations) in $\E^d$. The transitive action of $g
\in \cal G$ on $K \in \cal K$ is denoted by $gK$. Then
$$
W_\alpha(gK)=W_\alpha(K)\;\;;\alpha=0 \ldots d \;\; , \eqno(13)
$$
i.e., the Minkowski functionals of a body are independent of its
location in space.
\medskip\noindent
$\underline{\rm Continuity:}$
If $K_n \rightarrow K$ for $n\rightarrow \infty$,
$K_n, K \in \cal K$ (with convergence defined
in terms of the Hausdorff metric for sets), then
$$
W_\alpha(K_n)\rightarrow W_\alpha(K)\;\;;\alpha=0 \ldots d\;\;.
\eqno(14)
$$
Intuitively, this continuity property expresses the fact that
an approximation of a convex body by convex polyhedra $K_n$, for
example, also yields an approximation of $W_{\alpha} (K)$ by
$W_{\alpha} (K_n)$.

\bigskip
So far we only considered the Minkowski functionals restricted to the
class $\cal K$ of convex
bodies. Since we want to apply these measures
to investigate the spatial pattern of coverage
models as described at the beginning of this section,
the definition of the Minkowski functionals must be extended to
non-convex sets such as ${\cal B}(r)$. This extension may
be achieved in the following way (Hadwiger 1957).

Let $\cal R$ denote the class of subsets of $\E^d$
which can be represented as a finite union of
sets from $\cal K$, i.e., $A \in \cal R$ if and only if $A=
\cup_{i=1}^N K_i$, $N < \infty$, $K_i \in \cal K$.
The class $\cal R$ also includes the empty set $\emptyset$. In the
first step, the Euler characteristic is introduced by
$$
\chi(A)=
\left\{
\matrix{
1&,& A \in {\cal K}\; , \; A \neq \emptyset \; , \cr
0&,&  A = \emptyset \; .
}\right. \eqno(15)
$$
and extended to $\cal R$ via additivity,
$$
\chi (A\cup B)=\chi(A)+\chi (B)-\chi (A\cap B) \eqno(16)
$$
for any $A,B \in \cal R$. In particular,
$$
\chi(\cup_{i=1}^NK_i)=
\sum_i \chi(K_i)-\sum_{i<j} \chi(K_i \cap K_j)+ \ldots + (-1)^{N+1}
\chi(K_1\cap K_2 \cap \ldots \cap K_N)\; , \eqno(17)
$$
which follows from eq. (12) by induction.
The right hand side of eq. (17) only involves convex
sets and may be applied together with eq. (15)
to compute $\chi(A)$ for any $A \in \cal R$, as
illustrated in Fig.2. We also note that
$\chi : {\cal R} \rightarrow \Z$ is motion
invariant and it can be shown to agree with the Euler characteristic
as defined in algebraic
topology. In the cosmology literature (e.g. Melott 1990)
the Euler characteristic is usually
associated with the surface $\partial A$ of a body $A$.
For closed $(d-1)$-dimensional surfaces
in $\E^d$ one has the simple relationship
$$
\chi(\partial A)= \chi(A) \left(1+(-1)^{d-1}\right)\; .\eqno(18)
$$
\medskip\noindent
As a second step of the extension,
the Minkowski functionals are defined for $A\in \cal R$ by
$$
\matrix{
W_{\alpha}(A)&=&\int \chi(A \cap E_{\alpha})
d\mu(E_{\alpha})&,&\alpha =0,\ldots,d-1\;,\cr
 & & & &  \cr
W_{d}(A)&=&\omega_{d}\chi(A)\hfil&,
&\omega_{d}=\pi^{d/2}/\Gamma(1+d/2) \;.} \eqno(19)
$$
Here, $E_{\alpha}$ is a $\alpha-$dimensional plane in $\E^{d}$.
The integral runs over all positions
(induced by translations and rotations) of $E_\alpha$,
weighted with the so-called kinematical density
$d\mu(E_{\alpha})$ (Hadwiger 1957, Santal\'o 1976)
which is related to the invariant Haar
measure on the group of motions $\cal G$
and is normalized such that for a $d$-dimensional ball
$B_r$ with radius $r$,
$W_{\alpha}(B_r)=\omega_{d} r^{d-\alpha}$.
The volume of the unit ball ($B_{r=1}$) is $\omega_{d}$.

\medskip
In the case of convex $A \in \cal K$ the eqs. (19)
reproduce the Minkowski functionals obtained
previously from $V_\epsilon(A)$ (e.g. Miles 1974).
\medskip

According to their definition (19), the Minkowski functionals on $\cal
R$ inherit from the Euler characteristic the property of additivity,
i.e.,
$$
W_{\alpha} (\cup_{i=1}^N K_i ) = \sum_i W_{\alpha} (K_i)
- \sum_{i<j} W_{\alpha} (K_i \cap K_j) + \ldots + (-1)^{N+1}
W_{\alpha} (K_1 \cap \ldots \cap K_N )\;\;,\eqno(20)
$$
as well as motion invariance.
These features
together with their ``conditional continuity" (the $W_\alpha$ are
continuous when restricted to $\cal
K$) specify the Minkowski functionals
as a distinguished family of geometrical and
topological descriptors:
There is a remarkable theorem (Hadwiger 1957) which asserts that any
additive, motion invariant and conditionally continuous
functional $\cal F$ on subsets $A \subset \E^d \;,
\; A \in \cal R$, is a linear combination of
the $d+1$ Minkowski functionals,
$$
{\cal F}(A) = \sum_{\alpha=0}^{d} c_{\alpha}W_{\alpha}(A) \;\; ,
\eqno(21)
$$
with real coefficients $c_\alpha$ independent of $A$.
Important
consequences of this theorem are the ``principal kinematical formulae",
which may be written concisely in the form
$$
\int_{\cal G}M_{\alpha}(A \cap gB)dg=
\sum_{\beta=0}^{\alpha}{\alpha \choose
\beta}M_{\alpha-\beta}(B)M_{\beta}(A)
$$
$$ M_{\alpha}(A)={\omega_{d-\alpha} \over
\omega_\alpha \omega_d}W_{\alpha}(A)
\;,\;\;\alpha=0,\ldots ,d \; .  \eqno(22)
$$
The integral is performed with respect
to the invariant Haar measure $dg$ of $\cal G$ and runs
over all motions of the set $B$, with $A,B \in \cal R$.
In the case $B=B_\epsilon (x)$ and
$A=K \in \cal K$ the kinematical formula for $\alpha=d$
reproduces Steiner's formula (7),
$$
\int_{\R^d} \chi(K \cap B_\epsilon (x))d^dx=V_\epsilon(K)\;. \eqno(23)
$$

The kinematic formulae are useful mathematical tools in stereology
and stochastic geometry.
The Minkowski functionals can be calculated efficiently
for any given coverage without requiring
statistical assumptions about the underlying point set.
However, when the latter constitutes a
realization of a random spatial process (which is the case in
cosmological models for structure formation) then one faces a difficult
problem:
to infer the statistics of random geometrical patterns from a generally
limited amount of point set data. It
turns out that the mean values of the Minkowski functionals
provide {\it unbiased stereological
estimators} (Weil 1983).
Moreover, these mean values can be calculated exactly for the classical
Boolean model (Mecke and Wagner 1991), where the centers
of balls are distributed in $\E^d$
according to Poisson's law, which is often employed
as a reference model.

\noindent
For convenience, we recall the result for these mean values
per ball $B_r$ and for $d=3$:
$$
\eqalignno{
&<{V\over N}> = {1 \over \rho} (1-e^{-\nu}) \;\;,&(24a)\cr
&<{F\over N}> = 4 \pi r^2 e^{-\nu} \;\;, &(24b)\cr
&<{H\over N}> = 4 \pi r (1-{3\pi^2 \over 32} \nu) e^{-\nu}\;\;,&(24c)\cr
&<{G\over N}> = 4 \pi <\chi>=4 \pi (1-3\nu+{3\pi^2 \over 32} \nu^2)
e^{-\nu}\;\;, &(24d)\cr}
$$
with $\nu=4\pi r^3 \rho /3$ and the number density $\rho$ of balls.

\medskip

In the general case where the statistics of a point set in a domain
$\subset \E^3$ with volume $\Omega$ is specified by a sequence
of density correlation functions $\lbrace \rho_n (x_1, \ldots, x_n)
\vert n= 1,2,\ldots \rbrace$, the mean values for the augmented
coverage per unit volume are obtained from the additivity relation
(20) in the form
$$
< {1 \over \Omega} W_{\alpha} \lbrack {\cal B} (r) \rbrack >=
\sum_{n=1}^{\infty} { (-1)^{n+1} \over n{\rm !} \Omega}
\int \cdots \int_{\Omega} \rho_n (y_1, \ldots y_n) W_{\alpha}
\lbrack \cap_{i=1}^n B_r (y_i)\rbrack d^3 y_1 \ldots d^3 y_n \;\;.
\eqno(25)
$$
Obviously, the Minkowski functionals embody information from every order
$n$ of the correlations.

\vfill\eject
\noindent{\rmm 3. Examples}
\bigskip
\noindent
In this section we calculate the Minkowski functionals for
three examples of point processes in $\E^3$: a Poisson process, a
`double-Poisson' process and the ACO-catalogue of galaxy clusters.

\bigskip\smallskip
\noindent{\rma 3.1. Calculation method}
\medskip
\noindent
The Minkowski functionals $W_{\alpha}\lbrack {\cal B}(r) \rbrack$
for a covering ${\cal B}(r)=\cup_{i=1}^N B_r (x_i)$ of a given point
set $\lbrace x_1, \ldots, x_N \rbrace \subset \E^3$ may be calculated
straightforwardly via the additivity relation (20). However, in the
cases $\alpha \ge 1$ this algorithm becomes inefficient when the amount
of overlap between the augmented balls is excessive, since one has to
compute many redundant and mutually cancelling terms.
Therefore, we proceed, for $\alpha \ge 1$, alternatively as follows
(Mecke 1993):

\smallskip\noindent
The covering ${\cal B}(r)$ divides the whole space by piecewise smooth
spherical contours. These pieces are joined along circular arcs
where two balls intersect, and at singular points (vertices) which have
equal distance to the centers of three balls. Configurations where four
or more arcs have a point in common carry negligible weight and can be
ignored. Consider a ball $B_r (x_i)$ which is only partially covered
by other balls. The area $F_i$ and the integral curvatures $H_i$ and
$G_i$ are given by
$$
F_i = \int_{S_i} df = 3 W_1 (S_i)\;\;,\;\;
H_i = {1\over r}F_i = 3 W_2 (S_i)\;\;,\;\;
G_i = {1\over r^2}F_i = 3 W_3 (S_i)\;\;.\eqno(26)
$$
An uncovered arc $A_{ij}$ which length $\ell_{ij}$ on the intersection
$B_r (x_i) \cap B_r (x_j)$, $\Vert x_i - x_j \Vert = a_{ij} < 2r$,
gives rise to the curvature contributions
$$
H_{ij} = {1\over 2} \alpha_{ij} \ell_{ij} = 3 W_2 (A_{ij})\;\;\;,\;\;\;
G_{ij} = {a_{ij} \ell_{ij}\over r \sqrt{r^2 - (a_{ij}/2)^2}}
= 3 W_3 (A_{ij})\;\;,\eqno(27)
$$
where $\alpha_{ij}$ denotes the angle between the normals of
$B_r (x_i)$ and $B_r (x_j)$ along $A_{ij}$.

\noindent
The Gaussian curvature $G_{ijk} = 3 W_3 (P_{ijk})$ at an uncovered
vertex $P_{ijk}$ on the common intersection of three balls
centered at $x_i, x_j, x_k$ equals the solid angle spanned by the
three normals of the balls at $P_{ijk}$ and is obtained from
l'Huilier's formula
$$
\lbrack \tan ({1\over 4}G_{ijk}) \rbrack^2 =
$$
$$
\tan ({\alpha_1 + \alpha_2 + \alpha_3 \over 4})
\tan ({\alpha_1 + \alpha_2 - \alpha_3 \over 4})
\tan ({\alpha_1 - \alpha_2 + \alpha_3 \over 4})
\tan ({-\alpha_1 + \alpha_2 + \alpha_3 \over 4})\;,
$$
$$
\sin ({\alpha_1 \over 2}) ={\Vert x_i - x_j \Vert \over 2r}\;,\;
\sin ({\alpha_2 \over 2}) ={\Vert x_i - x_k \Vert \over 2r}\;,\;
\sin ({\alpha_3 \over 2}) ={\Vert x_j - x_k \Vert \over 2r}\;.\eqno(28)
$$
Finally, the total values of $W_{\alpha} \lbrack {\cal B}(r) \rbrack$
are given by
$$
W_1 \lbrack {\cal B}(r) \rbrack = \sum_i W_1 (S_i) \;\;,
$$
$$
W_2 \lbrack {\cal B}(r) \rbrack = \sum_i W_2 (S_i) - \sum_{ij} W_2
(A_{ij})\;\;,
$$
$$
W_3 \lbrack {\cal B}(r) \rbrack = \sum_i W_3 (S_i) - \sum_{ij} W_3
(A_{ij}) + \sum_{ijk} W_3 (P_{ijk}) = 4\pi \chi \lbrack {\cal
B}(r) \rbrack \;\;,\eqno(29)
$$
where the sums run over all uncovered surface pieces, arcs and vertices.

\medskip

In the figures below we display the reduced values of the Minkowski
functionals $\Phi_{\alpha}=W_{\alpha}\lbrack {\cal B}(r) \rbrack /
N W_{\alpha} (B_r)$, where $N$ is the number of sample points, and
$W_{\alpha} (B_r)={4\pi\over 3} r^{d-\alpha},d=3$; in particular,
$\Phi_3 = \chi \lbrack {\cal B}(r)\rbrack / N$.

As a consequence of the additivity of Minkowski functionals,
the statistical errors are always
smaller than the graphical symbols used in the figures.
The computation time on a local PC requires
a few CPU-seconds for a sample of
hundred points, increasing with $N\log(N)$ with the number $N$
of points.

\medskip

In the equations (29) the global Minkowski functionals are expressed in
terms of {\it local} measures. The latter also allow us to
introduce individual Minkowski functionals $W_{\alpha} (x_i;r)$ for
each ball $B_r (x_i)$($=:B_i$, for short).
in the following way:
$$
W_{\alpha} (x_i;r) = W_{\alpha} (S_i) - {1\over 2} \sum_{j,j \ne i}
W_{\alpha} (A_{ij}) \chi (B_i \cap B_j) + {1\over 3}
\sum_{j<k,j,k \ne i} W_{\alpha} (P_{ijk}) \chi (B_i \cap B_j \cap B_k)
\;\;.\eqno(30)
$$
The Euler characteristics in equation (30) equal unity if the convex
intersections are non-empty and vanish otherwise.
The factors $1/2$ and $1/3$ arise, since an arc and a vertex are shared
by two and three balls, respectively. Moreover,
$W_1 (A_{ij}) = W_1 (P_{ijk}) = W_2 (P_{ijk}) = 0$.
These individual local measures (30) provide means for defining
subsamples by selecting for fixed $r$ those balls whose associated
measures attain values in a prescribed interval. Examples are shown in
Fig.6 below.

\vfill\eject
\noindent{\rma 3.2. Poisson process}
\medskip
\noindent
In the first example the coverage ${\cal B}(r)$ is based on $10^4$
points, independently distributed in a unit cube with number density
$\rho$. We use the length scale $x = \rho^{1/3} r$,
normalized by the mean distance of the points; thus, the edge-length
of the cube is $x_{\rm cube}=21.6$. The radius of the decorating balls
is varied within the interval $0 \le x \le 1.6$; its maximum value
is small enough compared with $x_{\rm cube}$ to avoid
finite-size effects.
Fig.3a displays the values of $\Phi_{\alpha} (x), 1\le \alpha \le 3$.
The curves are the mean values $<\Phi_{\alpha} (x)>$ as obtained
from eqs. (24). For small radii, ${\cal B}(r)$ consists of isolated
balls with negligible overlap; therefore, each measure starts out at
unity. As the overlap increases, the measures decrease together with
the reduced total surface area $\Phi_1 (x)$ of the coverage.
$\Phi_2 (x)$ changes its sign, since the curvature of the singular edges
at intersections is negative and starts to dominate the posi\-tive
contribution from the spherical parts of the surface as the overlap
increases. Negative values of the Euler characteristic $\Phi_3 (x)$
are typical for a highly connected, ``sponge-like'' structure with
many tunnels. We also note that the first zero of $\Phi_3 (x)$
at $x_p \approx  0.44$ provides an estimate for the percolation
threshold (Mecke \& Wagner 1991).
\bigskip\smallskip
\bigskip
\noindent{\rma 3.3. `Double-Poisson' process}
\medskip
\noindent
This process was implemented as follows:
$50$ centers of spherical shells were randomly placed in a unit cube.
Then, $200$ points were distributed independently within each shell.
The resulting total number of points in the cube turned out to be
$6841$. The edge-length of the cube was $x_{\rm cube}=19.0$,
the inner and the outer radius of each shell has been chosen as
$x_i = 3.4$ and $x_o = 3.8$.

Fig.4 presents the projections of $750$ points from a thin planar
slice cut out from the cube.
For comparison, we also show a two-dimensional Poisson process with
$500$ points.

The measures $\Phi_{\alpha} (x)$ for the
`double-Poisson' process are given in Fig.3b. In comparison with Fig.3a,
the $x-$range with negative Euler characteristic is extended:
As $x$ increases, the balls start to overlap earlier and they fill the
space later. This feature is also reflected in the filaments
and bigger voids recognizable in Fig.4. The minimum value of the Euler
characteristic is higher than in the pure Poisson case, indicating that
now the structure is less ``spongy'' with fewer (but bigger) tunnels.

\vfill\eject
\noindent{\rma 3.4. ACO-catalogue}
\medskip
\noindent
For the third example of a point process we take
the ACO-catalogue of galaxy clusters (Abell \etal 1989, see
also Postman
\etal 1992 and the recent paper on the topological
characterization of the catalogue by Rhoads \etal 1993).
We consider a spherical section of the northern hemisphere with an
opening angle of $120^o$ and a radius of $450$\hn Mpc
or $z=0.15$. Each cluster in this section is taken into account, but
we neglect thinning in distance and other
possible deviations from a `fair sample'. In this way we obtain
$923$ clusters of galaxies;
the radius of the section is
$x_{\rm section}=9.6$ in normalized units, again
big enough to avoid edge effects, as in the previous
examples. Since the clusters are already extended objects, the ball
radii are chosen such that
$x\ge 0.06$ ($r_{\rm cluster} \ge 3$\hn Mpc).
In Fig.3c the Minkowski measures for the coverage of
this observed point process are plotted.

To exhibit the significance of these measures we display in Fig.5
the data for the Euler characteristic per unit surface area
$\chi^{*}(x):=\chi (x)/ F(x)$ for the three examples.
The enhancement of $\chi^{*} (x)$ at larger radii results from the
reduction of the size of uncovered tunnels. The similarity between the
data obtained from the ACO-catalogue and those from the `double-Poisson'
process is remarkable in view of the fact that the parameters chosen in
the latter have not been adjusted or optimized.

\medskip

In Fig.6 we show planar projections of three-dimensional ACO--subsamples
selected by the values of the individual Euler characteristics
of partially covered balls, as described in Section 3.1.
Only the centers of the balls are displayed.

\vfill\eject

\noindent{\rmm 4. Concluding Remarks}
\bigskip
\noindent
We have presented a method to analyze the spatial pattern of
galaxy distributions by employing the Minkowski functionals
which provide a complete family of morphological measures.
In order to apply this concept to the point-set representing
a sample of galaxies, we introduced a coverage by marking each member of
the point-set with the center of a ball.
The common radius of the balls is used as a diagnostic parameter for
displaying the morphological features of the coverage on different
length-scales, i.e., with varying spatial resolution. In
three-dimensional space the family of Minkowski measures consists of
the Euler characteristic, the covered volume, the surface area of the
coverage, and its integral mean curvature.

\noindent
The important aspects of the method may be summarized as follows.
\medskip\noindent
$\bullet$ On the practical side, the calculation of the Minkowski
functionals is {\it precise} and {\it CPU time effective}. In contrast
to the currently popular approach (see, e.g., Melott 1990) for
measuring the Euler characteristic (or the genus $g=1-\chi$) of
iso-density contours which requires the choice of two parameters,
namely a smoothing length and a density level, the present method only
involves the radius of the covering balls as a {\it single} scale.

\medskip\noindent
$\bullet$ As a consequence of their additivity, the Minkowski
functionals are {\it robust} and {\it global} measures: They yield
stable results even for small samples.

\medskip\noindent
$\bullet$ The method is efficient in {\it discriminating} different
spatial patterns and theoretical models.
Moreover, the method does not depend on assumptions about the statistics
of patterns, or model realizations, respectively.

\medskip\noindent
$\bullet$ The {\it mean} values of the Minkowski functionals provide
{\it statistically unbiased} descriptors which contain features of
$n$--point correlation functions at {\it any} order $n$.
Therefore, the present method is complementary to the conventional
approach restricted to lower-order correlation functions.

\medskip
Finally, the stereological properties of Minkowski functionals
(see, e.g., Weil 1983) offer means to infer morphological
characteristics of three-dimensional samples from lower-dimensional
sections such as ``pencil beams'' or two-dimensional all--sky surveys.
These stereological applications as well as a more detailed statistical
analysis is postponed to future work.
\bigskip
\noindent
{\rma Acknowledgements:} {\baselineskip=13pt
\mfs We would like to thank John A. Peacock
for providing us the new Abell cluster catalogue, and J\"urgen Ehlers
for valuable remarks on the manuscript.
TB is supported by DFG (Deutsche Forschungsgemeinschaft).}
\vfill\eject

\baselineskip=14.5pt

\centerline{\rmm References}
\bigskip\bigskip
{\mfs
\citem
Abell G.O., Corwin H.G., Olowin R.P. (1989): {\it Ap.J.Suppl.} {\bf 70},
1.
\citem
Bahcall N.A., Soneira R.M. (1983): {\it Ap.J.} {\bf 270}, 20.
\citem
Barrow J.D., Bhavsar S.P., Sonoda D.H. (1985): {\it M.N.R.A.S.} {\bf
216}, 17.
\citem
Brandenberger R.H., Kaplan D.M., Ramsey S.A. (1993): {\it M.N.R.A.S.},
submitted.
\citem
Buchert T., Mart\'\i nez V.J. (1993): {\it Ap.J.} {\bf 411}, 485.
\citem
Davis M., Peebles P.J.E. (1983): {\it Ap.J.} {\bf 267}, 465.
\citem
Efstathiou G., Kaiser N., Saunders W., Lawrence A., Rowan-Robinson M.,
Ellis R.S., Frenk C.S. (1990): {\it M.N.R.A.S.} {\bf 247}, 10P.
\citem
Geller M.J., Huchra J.P. (1989): {\it Science} {\bf 246}, 887.
\citem
Hadwiger H. (1957): {\it Vorlesungen \"uber Inhalt, Oberfl\"ache und
Isoperimetrie}, Springer, (in German).
\citem
Icke V., van de Weygaert R. (1987): {\it Astron. Astrophys.} {\bf
184}, 16.
\citem
Mart\'\i nez V.J., Jones B.J.T., Dominguez-Tenreiro R., van de
Weygaert R. (1990): {\it Ap.J.} {\bf 357}, 50.
\citem
Mart\'\i nez V.J. (1991): in: {\it Applying Fractals to Astronomy},
Lecture Notes in Physics, eds.: A. Heck, J. Perdang, Springer,
pp. 135-159.
\citem
Mart\'\i nez V.J., Portilla M., Jones B.J.T., Paredes S.
(1993): {\it Astron. Astrophys.}, in press.
\citem
Mecke K.R., Wagner H. (1991): {\it J. Stat. Phys.} {\bf 64}, 843.
\citem
Mecke K.R. (1993): {\it Ph.D.-Thesis}, University of Munich (in
German).
\citem
Melott A.L. (1988): {\it G.R.G.} {\bf 21}, 495.
\citem
Melott A.L. (1990): {\it Phys. Reports} {\bf 193}, 1.
\citem
Miles R.E. (1975): {\it Adv. Appl. Prob.} {\bf 7}, 818.
\citem
Mo H.J., Buchert T. (1990): {\it Astron. Astrophys.} {\bf 234}, 5.
\citem
Mo H.J., Deng Z.G., Xia X.Y., Schiller P., B\"orner G. (1992): {\it
Astron. Astrophys.} {\bf 257}, 1.
\citem
Peebles P.J.E. (1980): {\it The Large-scale Structure of the Universe},
Princeton Univ. Press.
\citem
Postman M., Huchra J.P., Geller M.J. (1992): {\it Ap.J} {\bf 384},
404.
\citem
Rhoads J.E., Gott III J.R., Postman M. (1993): {\it Ap.J.}, in press.
\citem
Rosenfeld A., Kak A.C. (1976): {\it Digital Picture Processing},
Academic Press.
\citem
Santal\'o L.A. (1976):
{\it Integral Geometry and Geometric Probability}, Addison-Wesley.
\citem
Serra J. (1982): {\it Image Analysis and Mathematical Morphology} {\bf
Vol.1,2}, Academic Press.
\citem
Shandarin S.F. (1983): {\it Sov. Astron. Lett.} {\bf 9}, 104.
\citem
van de Weygaert R. (1991): {\it Ph.D.-Thesis}, Leiden University.
\citem
Weil W. (1983): {\it Stereology}, in: {\it Convexity and its
Applications}, eds.: Gruber P.M., Wills J.M., Birkh\"auser.
\citem
White S.D.M. (1979): {\it M.N.R.A.S.} {\bf 186}, 145.
}

\vfill\eject
\baselineskip=15pt

\centerline{\rmm Figure Captions}
\bigskip\bigskip
{\mfs
\noindent{\bf Figure 1: } (a)
Parallel body of a convex polygone in $d=2$.
(b) The cylindrical segment
at the edge $k$ with dihedral angle $\alpha$
of a convex polyhedra contributes
$\epsilon \alpha L$ to the parallel volume;
therefore $3W_2(k)=\alpha L$.

\bigskip\smallskip

\noindent{\bf Figure 2:}
Illustrative examples
of the Euler characteristic
in two and three dimensions.
In two dimensions $\chi$ is the difference of components and holes,
whereas in three dimensions
it is the sum of the disconnected components
and cavities minus the number of tunnels. A negative
Euler characteristic indicates network-like structures.
Bottom: Illustrative calculation of $\chi$ using additivity.

\bigskip\smallskip

\noindent{\bf Figure 3:} Mean values of the Minkowski functionals,
i.e., of the reduced surface area
$\Phi_1 (x)$ (full rectangles), the reduced
integral mean curvature $\Phi_2 (x)$ (empty rectangles),
and the reduced
Euler characteristic $\Phi_3 (x)$ (triangles)
are depicted for a Poisson-process (3a), a
`double-Poisson' process (3b) and the ACO-catalogue (3c).
The curves in Fig.3a are obtained from the analytical formulas
(eqs. (24)).
The measures are normalized to 1 at $x=0$
and the radius $x$ of the balls centered at each
point of the process is measured in units of the mean
distance $\rho^{-1/3}$ of the points, where
$\rho$ is the number density of the process.

\bigskip\smallskip

\noindent{\bf Figure 4:}
Thin planar slices from
a three-dimensional cube: the upper panel
shows $500$ points which are independently distributed
(Poisson-process); the lower panel shows $750$ points which
are distributed according to
the `double-Poisson' process as explained in Section 3.3.

\bigskip\smallskip

\noindent{\bf Figure 5:}
The mean Euler characteristic per surface area $\chi^{*}(x)$
for the three point processes discussed:
a Poisson-process (P), a
'double-Poisson' process (DP), and the ACO-catalogue (ACO).
The zeros near $x_p=0.44$ indicate similar percolation thresholds.
The discrepancy of (P) and (ACO) obviously rules out the
Poisson-process as a statistical model, whereas the similarity
of (DP) and (ACO) is remarkable.

\bigskip\smallskip

\noindent{\bf Figure 6:} Subsamples from the
ACO-catalogue with the individual Euler characteristic
$\chi$ of each ball (with reduced radius $x$) as selection criterion,
projected onto a plane.
This illustrates the possibility
to classify
cluster environments by using the concept of {\it local morphology}
introduced in this work.

}

\vfill\eject
\bye